\documentclass{article}

\usepackage{arxiv}

\usepackage[utf8]{inputenc} 
\usepackage[T1]{fontenc}    
\usepackage{hyperref}       
\usepackage{url}            
\usepackage{booktabs}       
\usepackage{amsfonts}       
\usepackage{nicefrac}       
\usepackage{microtype}      
\usepackage{lipsum}
\usepackage[pdftex]{graphicx} 
\usepackage{amsmath}
\usepackage{graphicx}
\usepackage[section]{placeins}
\usepackage{multirow}
\usepackage{multicol}
\usepackage{algorithm}
\usepackage{algorithmic}

\usepackage[square,numbers]{natbib}
\title{Defending Against Intelligent Attackers at Large Scales}

\author{
  Andrew J. Lohn\\
  Center for Security and Emerging Technology, Georgetown University\\
  Washington, DC, USA \\
  \texttt{drew.lohn@georgetown.edu} \\
}

\begin{document}
\maketitle

\begin{abstract}
We investigate the scale of attack and defense mathematically in the context of AI's possible effect on cybersecurity. For a given target today, highly scaled cyber attacks such as from worms or botnets typically all fail or all succeed. Here, we consider the effect of scale if those attack agents were intelligent and creative enough to act independently such that each attack attempt was different from the others or such that attackers could learn from their successes and failures. We find that small increases in the number or quality of defenses can compensate for exponential increases in the number of independent attacks and for exponential speedups.  
\end{abstract}


\section{Introduction}
\label{sec:introduction}
\subsection{The Scale of Cyber Attack and Defense}
Despite automation in offensive cyber such as with botnets or worms, and in defensive cyber such as with malware scanners, anomaly detectors, and packet filters, both attack and defense are still largely human-directed. As AI promises to increase the speed and scale of both offense and defense, the obvious question is which side benefits more.\cite{lohnODbalance,corsi2024considerations,tang2024implications} Will AI substantially increase the scale or speed of cyber attack or defense. 

The challenge is not simply scale or speed. Worms, bots, and shared malware all scale extremely well but are not very independent.\cite{botnets} If an attack agent tries the same attack technique, a target's defenses will typically hold against all of the attack attempts if they hold against one. That is somewhat less true in attacks such as DDoS, where the victim is not being penetrated or manipulated, just overwhelmed.\cite{DDoS} But even for DDoS, there are layers of filters and defenses.\cite{DDoSdefense} For other types of intrusions, the attack attempts would need to be more varied or independent to overcome a defense that does not fold initially. AI promises to provide that originality while also promising to scale the number of those independent attack attempts.\cite{deepmindCyberAIFramwork}

Defenders already face a variety of creative human attackers and do not rely on a single defense. In some cases, such as an exposed web service where a SQL injection or heartbleed attack interfaces directly, there may be fewer lines of defense.\cite{heartbleed} Examples of defenses in that case may be a web app firewall or input validation. For more enterprise defenses such as those that protect corporate intellectual property or critical infrastructure, there are many possible defenses, vendors, and products that can be used. There are passwords, firewalls, antivirus, anomaly detection, intrusion detection, restricted permissions, etc. AI promises to increase the number of products and services that defenders can purchase or that they can effectively monitor. And AI assistance may allow defenders to harden those defenses such as by setting more aggressive thresholds for alerts if AI can make inspecting false positives less onerous. 

\subsection{Defense in Depth}

Defense in depth can be developed according to at least two different strategies.\cite{DinD} A Blockade strategy is where a sufficient number of defenses are built that attackers are unable to penentrate them all even if those defenses are individually weak. An alternative strategy does not require any of the defenses to hold. The Delay strategy uses each defense to slow the attack enough that defenders can discover and remove attackers before all defenses are overcome. That delay can be considered in terms of how many seconds or months does a defense slow attackers or in terms of how many attempts do attackers need in order to breach their target.

We have previously developed simple mathematical representations of these two strategies.\cite{DinD} Here we evaluate the effect of scaling up the number of attacks and defenses. We also develop a combine approach for the first time where a delay strategy is defined by number of attempts instead of time. Using these models, we hope to gain an intuition for how increasing the number of attacks and defenses will change the likelihood of success for attackers or defenders. That intuition is not meant to be a final answer for whether AI will offensively or defensively bias cyber operations. It is meant to provide a starting point for understanding the scaling dynamics and provide a framing for debate about those dynamics. Hopefully, it will illustrate the fundamental pressures that attackers and defenders will face with increasing scale and clarify the decisions each can make to either alleviate or increase those pressures.

\section{Blockade Strategies for Defense in Depth}
\label{sec:blockade}
Drawing from our previous work on defense in depth, equation \ref{blockade_equation} describes the likelihood, $L$, of at least one attacker success where all individual defenses are overcome. For this blockade strategy, the probability that an individual defense fails or is irrelevant is denoted by $p$, the number of individual defenses is $n$, and the number of attack attempts is $N$. 

A higher $p$ is worse for defenders and requires a larger $n$ to compensate. Hardness of individual defenses, or the probability that an individual defense will be effective, is $(1-p)$. We treat each defense as having the same value for $p$ as in an average. This is the conservative assumption because the weakest overall defense occurs when each individual defense has the same value of $p$ as follows from the arithmetic mean-geometric mean inequality. In the extreme, a two-defense system with $p=0$ and $p=1$ have an average $p=0.5$ but are certain to hold for every attack.

\begin{equation}
\label{blockade_equation}
L= 1-(1-p^n)^N 
\end{equation}

AI might allow defenders to implement or monitor more defenses, increasing $n$ in equation \ref{blockade_equation}. AI may also make existing defenses more effective. If AI can help investigate more alerts, then defenders can set more aggressive alert thresholds, increasing $p$. Alternatively, AI might invent better attack tactics than humans and increase $p$. 

\subsection{Number of Attacks}

One quantity of interest is how many additional defenses are needed to compensate for an increased number of attacks. We rearrange equation \ref{blockade_equation} to calculate the number of attacks required to overcome the defenses. 

\begin{equation}
\label{eqn_N_blockade}
    N = \frac{ln(1-L)}{ln(1-p^{n})}
\end{equation}

Equation \ref{eqn_N_blockade} can be simplified with a few approximations that are valid when both $L$ and $p^{n}$ are small. We approximate that $ln(1-L) \approx -L$ and that $ln(1-p^{n}) \approx -p^{n}$ in equation \ref{eqn_N_blockade_approx}.

\begin{equation}
\label{eqn_N_blockade_approx}
    N \approx \frac{L}{p^{n}}
\end{equation}

Equation \ref{eqn_N_blockade_approx} indicates that the number of attacks needed to compensate for an increasing number of defenses grows exponentially. It suggests that as attack and defense scale, small increases in the number of independent defenses can repel exponentially increasing numbers of independent attacks. That could be important from a cyber perspective because defenses may be difficult to scale to large numbers while maintaining productivity.\cite{usability} It may be difficult to implement many defenses if each defense interferes somewhat with usability or productivity such as by slowing digital processes, annoying users, or adding expense.

\begin{figure}[h]
\centering
\includegraphics[width=0.8\textwidth]{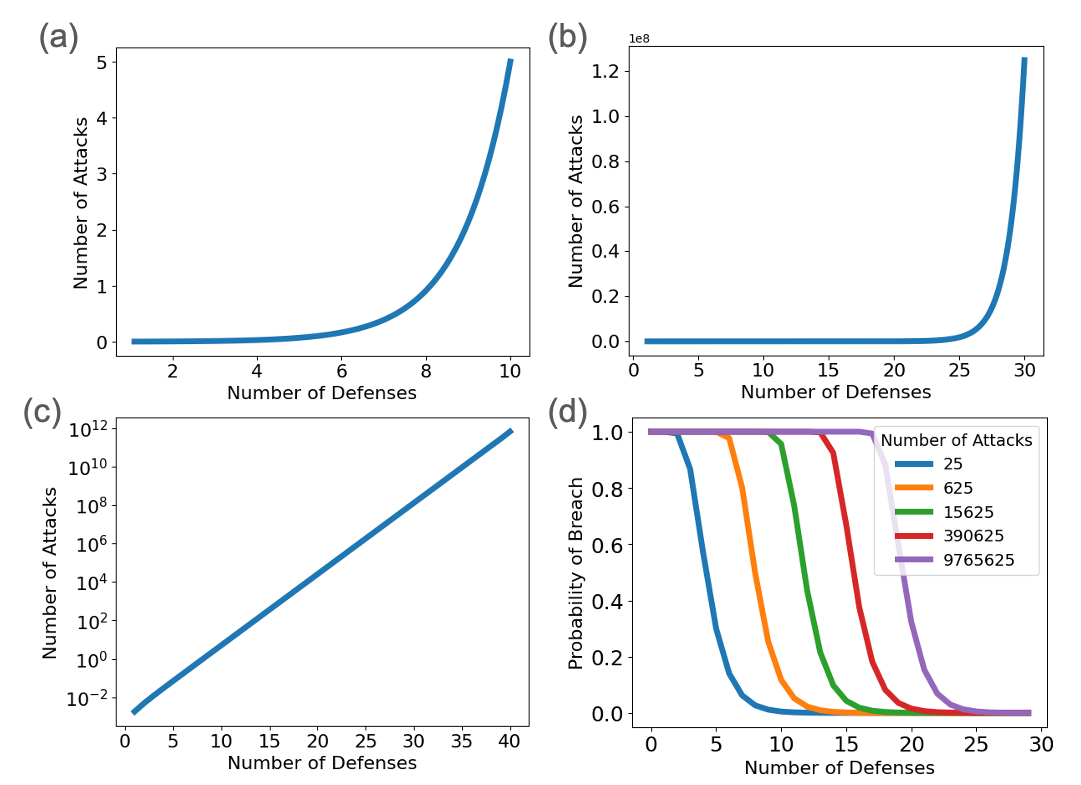}
\caption{The number of independent attacks that would be needed to overcome a set of defenses increases exponentially with the number of defenses. Parts a, b, and c are for a constant probability (0.001) that at least one attack succeeds against all defenses. Part d shows the likelihood of at least one breach while holding the probability of individual defense failure ($p$) at 0.43. It shows linear spacing between exponentially increasing numbers of attacks.}
\label{figBlockadeN}
\end{figure}

Figure \ref{figBlockadeN} shows the exponential increase in number of attacks that is needed to keep the probability of no breaches constant. For parts a-c, the probability of at least one successful attack was held at 0.001. With just five attackers and ten defenses, that corresponds to individual defenses that fail 43\% of the time. But parts b and c show that even with such individually weak defenses, relatively small increases in the number of defenses can repel millions of independent attacks. Part d shows the same effect, plotting the probability of all defenses failing at least once, where each separate curve is for an exponentially increasing number of attacks.

\subsection{Hardness of Defenses}
\begin{figure}[h!]
\centering
\includegraphics[width=0.5\textwidth]{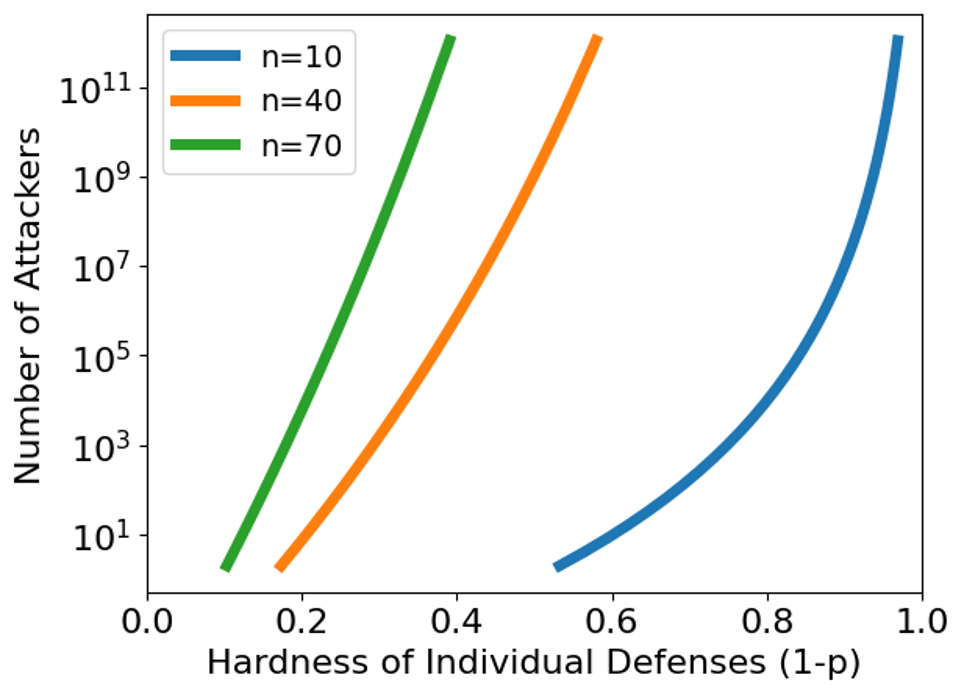}
\caption{Harder individual defenses can compensate for exponentially increasing numbers of attacks. The probability of all defenses failing at least once, $L$, was held constant at 0.001 in the figure.}
\label{figBlockadeP}
\end{figure}

Rather than adjusting the number of defenses, defenders may seek to increase the hardness of their defenses or to understand the effect of weakened defenses. Rearranging equation \ref{blockade_equation}, we calculate how hard individual defenses must be to compensate for an increasing scale of attacks in equation \ref{pNrates}. Equation \ref{pNrates} is illustrated in figure \ref{figBlockadeP} where the likelihood of at least one breach was again held constat at 0.001. Sublinear increases in hardness are able to repel exponentially increasing numbers of attacks, and smaller increases in individual hardness are required for larger the numbers of defenses. By the same logic, weakening defenses has the same effect as exponentially increasing the number of attacks.

\begin{equation}
\label{pNrates}
    p = (1 -(1-L)^{1/N})^{1/n}
\end{equation}

\section{Delay Strategies for Defense in Depth}
\label{sec:delay}
A Delay strategy for defense in depth does not rely on any of the individual defenses to hold. The individual defenses are only intended to impose a delay on attackers and alert the defender who can then discover and remove attackers before all defenses are overcome. Following our previous work that assumes a constant rate of discovery, the probability of any individual attack overcoming all defenses is $L_{I}$ described in equation \ref{delay_individuals}, where $\lambda$ is the rate of detection at each individual defense and $\tau$ is the time for attackers to overcome a defense.\cite{DinD} The likelihood of at least one attack succeeding ($L$), number of attacks ($N$), and number of attackers ($n$) all have the same meaning as in the Blockade section.

\begin{subequations}
\label{eqnDelay}
\begin{gather}
    \label{delay_individuals}
    L_I = \prod_{i=1}^{n} e^{-\lambda \tau} = e^{-\lambda \tau n} \\
    \label{delay_aggregate}
    L = 1- (1-e^{-\lambda \tau n})^N
\end{gather}
\end{subequations}

Equation \ref{delay_aggregate} is similar to equation \ref{blockade_equation} from the previous section. Where the previous section had individual defense failure rates ($p$), that part of the equation is replaced by the likelihood of a discovery at each defense ($e^{-\lambda \tau}$) here. As a result, many of the dynamics of Delay strategies are similar to those in Blockade strategies.

\subsection{Speedups and Number of Attacks}

Equation \ref{delay_individuals} suggests that increasing the speed of both attack ($\tau^{-1}$) and defense ($\lambda$) by the same factor ($s$) has no effect, but that ignores that a speedup could increase the number of attacks ($N$). The total number of attacks can be calculated by multiplying the number of separate attackers ($N_{a}$) by the number of sequential attacks that each attacker can attempt over a period of time $T$. The expected amount of time per attack is the characteristic time to overcome a defense ($\tau$) multiplied by the expected number of defenses overcome, $E[n_{overcome}] =\frac{1}{1-p} = \frac{1}{e^{-\lambda \tau}}$. The resulting likelihood of at least one breach over a given period of time is shown in equation \ref{eqnLT}.

\begin{subequations}
\begin{gather}
    N = N_{a} \frac{Tse^{-\lambda \tau}}{\tau} \\
    \label{eqnLT}
    L = 1 - (1-e^{-\lambda \tau n})^{N_{a}Ts\tau^{-1}e^{-\lambda \tau}}
\end{gather}
\end{subequations}

Again, small increases in the number of defenses ($n$) can compensate for large increases in the number of attacks ($N$), regardless of whether those are the result of more attackers ($N_{a}$) or faster attacks ($s$). We see that by comparing the increase in speed ($s$) to a compensating increase in number of defenses ($n$) in equation \ref{s_N} which makes use of the approximations $ln(1-L) \approx -L$ and $ln(1-e^{-\lambda \tau n}) \approx -e^{-\lambda \tau n}$.

\begin{equation}
    \label{s_N}
    s \approx \frac{L \tau}{N_{a}T}e^{\lambda \tau (n+1)}
\end{equation}

\begin{figure}[h!]
\centering
\includegraphics[width=0.8\textwidth]{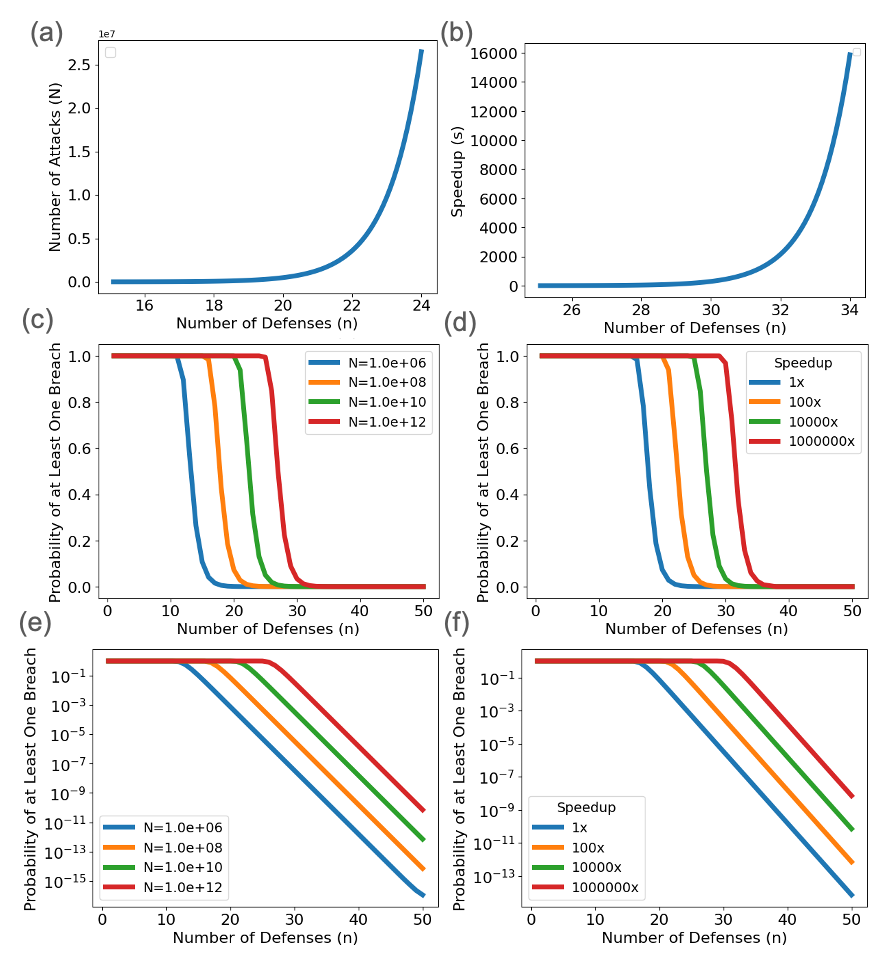}
\caption{An exponential increase in the number of attacks or speedups can be compensated for by a linear increase in the number of defenses. For parts a and b, the likelihood of at least one breach ($L$) is held constant at 0.001.}
\label{figDelayNumberDefenses}
\end{figure}

Again, an exponential increase in the threat from speedups or increased number of attacks can be compensated for by small linear increases in the number of defenses. Figure \ref{figDelayNumberDefenses} illustrates those relations. 

In parts a and b of figure \ref{figDelayNumberDefenses}, the likelihood of breach is held at 0.001 and the ratio of defensive speed to offensive speed ($\lambda \tau$) is set to one. For part b, the number of attackers ($N_{a}$) was 1,000 and the period of attack ($T$) was set to 100,000 times the characteristic time to defeat an individual defense ($\tau$).

Parts c and e show the same graph on linear and semilog scales respectively, where an exponentially increasing number of attacks is compensated by small linear increases in number of defenses. About twenty defenses are needed to defend against millions of attacks, but not so many more are needed to defend against trillions. The same effect is shown in parts d and f for speedups on linear and semilog axes respectively. A million times speedup requires only a relatively small increase in the number of defenses.

\subsection{Differential Speedups Between Offense and Defense}

\begin{figure}[h!]
\centering
\includegraphics[width=0.8\textwidth]{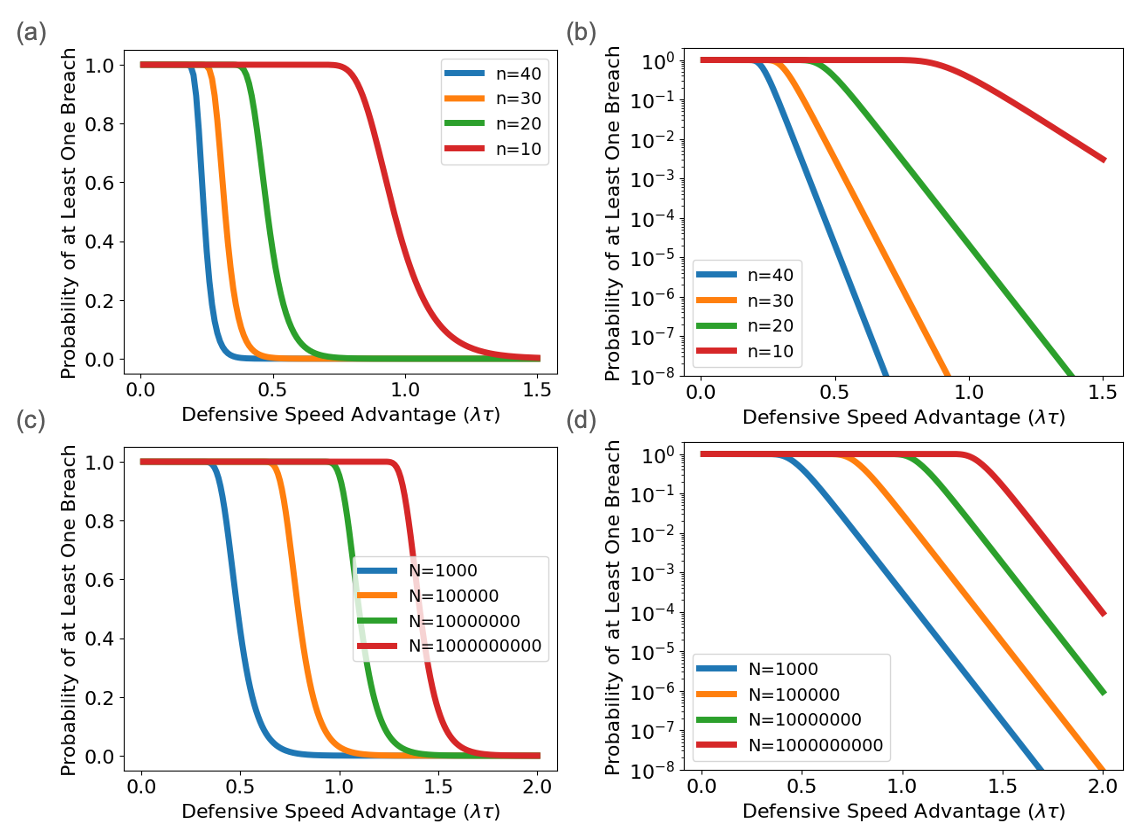}
\caption{The number of defenses and the number of attacks are sensitive to the ratio of defensive and offensive rates detection and defense bypass respectively.}
\label{figDelayRatioRates}
\end{figure}

The likelihood of at least one breach is more sensitive to changes in the relative speed of offense and defense. We refer to the the multiple $\lambda \tau$ as the defensive speed advantage where values less than one indicate a defensive disadvantage and values greater than one indicate that individual detections are typically fast compared to the rate of overcoming defenses. Figure \ref{figDelayRatioRates} shows the effect of small changes in that ratio. Parts a and b of figure \ref{figDelayRatioRates} show, on linear and semilog scales respectively, that the impact of any differences in defensive speed advantage are especially pronounced for small numbers of defenses. To compensate for a reduction in defensive speed advantage, the number of defenses ($n$) would need to increase linearly. By contrast, parts c and d show that small changes in defensive speed advantage correspond to exponential changes in number of attacks ($N$). 

It is hard to say what changes to expect in the relative speeds of offense and defense as a result of AI.\cite{speedups} Many tactics for both are already fast in ways that may be difficult to further accelerate. For example, directory or network scans are already automated, phishing for initial access or lateral movement is limited by the time for (primarily) human recipients to be victimized, files are transferred or exfiltrated at the bandwidth of the network, and harddrives encrypt at the same speed whether the command is given by a human or an AI. Defense has a few advantages with respect to speed though. Defense can choose to decelerate various processes such as by requiring human approvals before risky actions, throttling transfers, or account lock-outs.\cite{sludge} Further, attackers often choose to act slowly despite having the ability to act quickly in order to avoid triggering defensive alerts.\cite{lowAndSlow}

\section{Combined Blockade and Delay: Attackers That Learn}

So far we have assumed that AI advances to the point where attacks are independent; if one attack fails then others are not more likely to either succeed or fail. It is also worth also considering attacks that go beyond independent. In the extreme, we could assume that each defense will hold for the first probe but that the attacker will then know for certain how to defeat it in their second attempt. That individual defense could remain broken in all subsequent attack attempts. In that case, the number of attempts required would simply be $N=2n$.

Harder defenses may require a larger number of attempts to defeat an individual defense for the first time. The number of random trials to achieve the first success follows the geometric distribution, $(1-p)/p$. Treating that defense as continually broken from then on, the total number of attempts to defeat all defenses is calculated as the sum of geometric distributions. That is the negative binomial distribution shown in equation \ref{negBinDist}.

\begin{equation}
    \label{negBinDist}
    N = n(1-p)/p
\end{equation}

Again, we treat each individual defense as having the same hardness $(1-p)$. Doing so results in simpler math and is a conservative assumption for defenders, which follows from Jensen's inequality and from $(1-p)/p$ being a convex function.\cite{JensenInequality} The number of necessary attacks will always be greater than or equal to the number that is calculated by using the average of $p$ rather than individual values of $p$. As a tangible example, two defenses with $p=0.02$ result in $N=98$. Two defenses with $p=0.01$ and $p=0.03$ result in $N=131$.

Although equation \ref{negBinDist} is calculated using the probability that an individual defense will hold as in a Blockade strategy, each defense will fail eventually and only imposes a delay on the attacker. That delay is counted in number of attempts rather than time directly. For this Delay strategy to be useful, the defender needs to have some chance of detecting the attacker. Defenders could have a probability of detecting an attacker in each attempt ($d$) or equivalently, a probability of going undetected $u=1-d$.

The probability of a single attacker being undetected in all of the attempts required to overcome all of the defenses is shown in \ref{eqnUndetectedSuccess}. Eqn \ref{eqnUndetectedSuccess} assumes that each defense could take many attempts to overcome and that each attempt is an opportunity for detection. Again, treating each $u$ as having the same value equal to the mean is a conservative assumption for defenders.

\begin{equation}
\label{eqnUndetectedSuccess}
    P_{N} = u^{n(1-p)/p} \\
\end{equation}

These calculations are for a single attacker's campaign but there could be many attackers. Each attacker could also re-engage after a campaign fails and they are detected and removed. For these calculations, we do not consider information sharing among attackers because attackers would presumably be reluctant to share too widely out of concern that defenders could listen in to learn that the defenses are broken and fix them. We also assume that when an attack is discovered, the discovery and defender's logs reveal which defenses were broken so that all defenses can be repaired such as by resetting passwords, reconfiguring permissions, writing new YARA rules, and patching software. 

Combining these concepts, the probability that at least one attack campaign overcomes all defenses without being detected is shown in equation \ref{eqnL_N_n}.

\begin{equation}
\label{eqnL_N_n}
    L_{N} = 1 - (1-u^{n(1-p)/p})^{N_{A}} \\
\end{equation}

It may be that the probability of an attack attempt going undetected ($u$) is high. In that case, many defenses may be required or the defenses may each need to be hard (small $p$) to force many attack attempts and create many opportunities for detection.

In these calculations for defending against multiple attackers, we have not considered that repairing defenses in response to one attacker may also repel other attackers. Forcing a password change or patching vulnerable software in response to one attacker would thwart all other attackers who discovered that password or that vulnerability but that effect is not considered in this model.

\subsection{Where to Improve Defenses}
When defending against a large number of attackers who can each learn from their previous attack attempts, defenders have several options. Defenders can try to increase the hardness of each defense (decrease $p$) so that more attempts are needed to break each one, they can increase the number of defenses (increase $n$), and they can increase the probability of detecting an attacker (decrease $u$, increase $d$). Ideally, defenders would do all of these simultaneously. In the case where defenses are hard (small $p$) and detection is weak (small $d$), viable defenses require a combination of these variables that satisfies relation \ref{rlnBD} which is derived in Appendix \ref{appendixA}. To have likelihoods of breach that are acceptable in practice, the left side of relation \ref{rlnBD} typically needs to be greater than one by more than a few, but not much greater than one.

\begin{equation}
\label{rlnBD}
    \frac{nd}{p} - ln(N_{A}) > 1
\end{equation}

Relations \ref{rlnBD} shows that improvements in any of $n$, $d$, or $p$ can each compensate for exponential increases in number of attackers ($N_{A}$). The defensive factors are also combined such that small improvements in each can be combined to compensate for large increases in number of attackers. 

As an example, one hundred thousand separate attackers would equate to a value of $ln(N_{A})$ of about 11.5. If the detection rates ($d$) and probability of individual defenses failing ($p$) are the same, then only about 15 separate defenses ($n$) would be needed to have a viable defense in depth. With 15 defenses that each require an average of 20 attempts to break, and a detection rate of 5\%, the likelihood of at least one breach from those hundred thousand attacks would be less than 5\%. Doubling any one of the number of defenses ($n$), the number of attempts required to break a defense ($1/p$), or the detection rate ($d$) would reduce the likelihood of breach dramatically, from $0.05$ to about $10^{-8}$. For comparison, simultaneously doubling the number of attackers, or equivalently, doubling the number of times that a single attacker can re-engage after being discovered, would result in a probability of about $2 \times 10^{-8}$.

\section{Conclusions}
\label{sec:conclusions}
If AI enables large scales of independent attacks, defenders may rely on defense in depth. Fortunately, it appears that layered defenses have some prospect of holding against such attacks so that relatively small increases in the number or quality of defenses may be able to protect against exponential increases in the number of independent attackers. These models are only simplifications of real systems but we hope that they help clarify the types of strategies or shortcomings for both attackers and defenders. We hope that the models can help design aggregate defenses that can succeed as AI increases the scale, speed, and originality of autonomous cyber attacks.

\section*{Acknowledgement}
The author would like to thank John Bansemer, Kendrea Beers, Colin Shea-Blymyer, Kyle Crichton, Rebecca Gelles, Jessica Ji, Kyle Miller, Neil Thompson, and Katherine Quinn for helpful critique.

\section*{Funding}
This work was supported by a grant from the United Kingdom’s Department for Science, Innovation, and Technology.

\bibliography{references}

\appendix
\section{Appendix A}
\label{appendixA}
Equation \ref{eqnA1} is the same as equation \ref{eqnL_N_n}. It describes the likelihood of at least one breach in a combined Blockade and Delay strategy. Once an attacker bypasses a defense, that defense remains broken in subsequent attempts by that attacker unless they are discovered. After an attacker is discovered, all of the defenses can be repaired. There can be many possible attackers ($N_{A}$), but each attacker must break defenses for themselves; a defense that is broken for one attacker can remain effective against another attacker. 

\begin{equation}
    \label{eqnA1}
    L = 1 - (1-u^{n(1-p)/p})^{N_{A}}
\end{equation}

We assume that each individual defense requires several attempts to break (small $p$), that detection rates are low (small $d=1-u$), and that there are many attackers (large $N_{A}$). For a set of defenses to be considered viable, the likelihood of a breach ($L$) must be small. We try to determine the combinations of $n$, $N_{A}$, $d$, and $p$ for which that is true. Rearranging equation \ref{eqnA1} slightly and taking the logarithm of both side:

\begin{subequations}
\begin{gather}
    ln(1-L) = N_{A}ln(1-u^{n(1-p)/p}) \\
    ln(1-L) = N_{A}ln(1-b), b=u^{n(1-p)/p}
\end{gather}
\end{subequations}

A small likelihood of breach implies that $ln(1-L)$ must be a negative number near zero. Equivalently, its inverse ($1/ln(1-L)$) must be a large negative number as described in relation \ref{eqnInverse}.

\begin{equation}
    \label{eqnInverse}
    \frac{-1}{N_{A}ln(1-b)} >> 1
\end{equation}

Relation \ref{eqnInverse} only holds if $b$ is small. In that case, $ln(1-b) \approx -b$ and relation \ref{eqnInverse} is approximately the following:

\begin{equation}
    \label{eqnInverseApprox}
    \frac{1}{bN_{A}} >> 1
\end{equation}

Taking the logarithm of both sides,

\begin{equation}
\label{eqnsumlogs}
    -ln(b) - ln(N_{A}) > 1
\end{equation}

We will now estimate $ln(b)$. In equation \ref{eqnlnb}, we make use of the definition that  $u=1-d$, the approximation that $(1-p) \approx 1$, and the approximation that $ln(1-d) \approx -d$.

\begin{subequations}
\label{eqnlnb}
\begin{gather}
    b = u^{n(1-p)/p} \approx u^{n/p} \\
    ln(b) \approx \frac{n}{p}ln(u) \approx \frac{-nd}{p} \\
\end{gather}
\end{subequations}

Substituting $ln(b)$ from equation \ref{eqnlnb} into equation \ref{eqnsumlogs}, we come to relation \ref{rlnBD}.

\begin{equation}
    \frac{nd}{p} - ln(N_{A}) > 1
\end{equation}

\end{document}